# Reinventing College Physics for Biologists: Explicating an epistemological curriculum

Edward F. Redish and David Hammer

*Departments of Physics and Curriculum & Instruction*
*University of Maryland, College Park, MD 20742-4111*

**Abstract.** The University of Maryland Physics Education Research Group (UMd-PERG) carried out a five-year research project to rethink, observe, and reform introductory algebra-based (college) physics. This class is one of the Maryland Physics Department's large service courses, serving primarily life-science majors. After consultation with biologists, we re-focused the class on helping the students learn to think scientifically – to build coherence, think in terms of mechanism, and to follow the implications of assumptions. We designed the course to tap into students' productive conceptual and epistemological resources, based on a theoretical framework from research on learning. The reformed class retains its traditional structure in terms of time and instructional personnel, but we modified existing best-practices curricular materials, including *Peer Instruction*, *Interactive Lecture Demonstrations*, and *Tutorials*. We provided class-controlled spaces for student collaboration, which allowed us to observe and record students learning directly. We also scanned all written homework and examinations, and we administered pre-post conceptual and epistemological surveys. The reformed class enhanced the strong gains on pre-post conceptual tests produced by the best-practices materials while obtaining unprecedented pre-post gains on epistemological surveys instead of the traditional losses.

**Keywords:** college physics, epistemology, implicit curriculum, physics for biologists, expectations.
**PACS:** 01.40.Di, 01.40.Fk

## RETHINKING ALGEBRA-BASED PHYSICS FOR BIOLOGISTS

Algebra-based (college) physics is one of most physics departments' largest service courses. At the University of Maryland, we teach approximately 800 students a year in each term of this two-semester class. Increasingly, the population is dominated by majors from our college of life sciences. These include many pre-health care majors, such as pre-medical, pre-dental, pre-physical therapy, pre-veterinary, etc., as well as a growing number of pre-research biologists.

In the years 2000-2005, the University of Maryland's Physics Education Research Group (UMd-PERG) carried out an NSF-supported research study to observe student behavior in algebra-based physics and to explore reforms in that class.[1] The reforms we created for the class were based on

- our reading of current writings on the needs of modern biology students,[2,3,4]
- interviews with biology faculty,
- a theoretical framework that gives us insight into how students think and learn about physics,[5,6,7] and
- our experiences in small seminar courses for college students and in high school courses.[8,9,10]

Every class not only contains its explicit content, but elements that are traditionally not made explicit in descriptions of the class – an *implicit curriculum*.[11] For example, traditional instructors tend to assume that students learn how to think about and do scientific reasoning while doing traditional class activities, such as reading the text and doing end of chapter problems. Some students do learn this successfully, but research indicates that most do not and indeed, some pick up bad habits and inappropriate modes of thinking.[12] We chose to focus the class on helping students learn how to learn science, content that is implicit in most courses and that the research convinced us needs to be addressed explicitly.

Many of these implicit elements are *epistemological* – issues about the nature of scientific knowledge: how we know what we know, how to create new knowledge via problem solving, how we make inferences, what makes sense, and how to build physical intuition. These issues have particular importance for the population of biologists currently dominating college physics classes, but they are equally important for other populations of physics students. The implicit epistemological content tacitly taught in traditional courses often is <u>not</u> what we want our students to learn; rather, it encourages poor approaches to learning such as rote memorization and the denigration of everyday experiences and intuitions.[13]

We transformed the class to encourage student learning to take place in class-managed areas where it could be observed and videotaped. We collected large amounts of written data, including pre-post conceptual and epistemological surveys, and digital scans of all written homework and exams. The instructor encouraged students to reflect briefly on the class in written essays. In addition, researchers in our group who were not part of the instructional team interviewed some students about their experiences in the class, pre and post. In this paper, we present the reforms we developed and review the broad evidence of their success.

We achieved what we believe to be the first documented large gains on an epistemological survey in a large lecture introductory physics class at a university; we did it while not only retaining but enhancing high values for the fractional gain on a mechanics conceptual survey; we produced large gains (compared to traditional class) on a split task post-instruction concept survey that measured not only students' knowledge of the correct results but their intuitive comfort with those results; and we documented in some detail the kinds of epistemological difficulties students run into during the class and the extent to which those difficulties can be overcome. All this was done within the context of a traditional environment with essentially the same resources provided to our standard large lecture class.

In section II of this paper, we describe our motivations for choosing the reforms we did. In section III we describe the theoretical base for our analysis both of our goals and the instructional tools we chose to reform. We describe the reforms we carried out in section IV. In section V we describe our methods for observing and evaluating the class, and we present our observations and conclusions in section VI. What we learned from detailed research studies that contribute to our understanding of how an individual learns physics are described in other publications.[14,15,16,17,18]

## DECIDING WHAT MATTERS FOR BIOSCIENCE STUDENTS

One of the most significant transformations in the landscape of science in the past half-century has been the growing strength of biology as a fundamental science. There is broad agreement among leading biology and medical researchers that future biology students will need to become much more knowledgeable in basic physics, chemistry, and math. It is not only a familiarity of the facts and vocabulary of those fields that these students require, but a deep understanding of the disciplinary patterns of knowledge and process including a solid understanding of scientific reasoning.

With the increasing role of physics and the pace of change in the biological professions, it is essential that physics education go beyond isolated facts and narrow procedures. More than helping students understand established ideas, science instruction must help them understand how those ideas, and further ideas we cannot now anticipate, come to be. Students must be prepared to contend with ambiguities, to make sound judgments about what to accept and what to question, to reconsider past assumptions and adapt to new discoveries. They must learn what a measurement means – and does not mean. They must learn how to evaluate their data and see its implications. In short, they must learn an *adaptive expertise* – the ability to respond effectively and productively to new situations and new knowledge as it develops.[19]

Science instruction at the university level tends to ignore these issues, hoping that they will somehow spontaneously spring into being through traditional coverage of traditional content. This appears to work only for a small minority of students after many years of combined undergraduate and graduate training. Our goal in this project was to learn how to help more students develop these broad thinking and learning skills by paying explicit attention to these issues and developing curriculum to deal with them.

## A RESOURCE-BASED MODEL OF MIND

Our redesign was based largely on a *resources* based view of students' knowledge and reasoning,[5,6,7] one that supports Einstein's famous claim that "The whole of science is nothing more than a refinement of everyday thinking."[20] Everyday thinking involves both *conceptual* and *epistemological* resources, and learning physics begins from marshalling those resources in productive ways.

Student conceptual resources include their extensive intuitive knowledge about physical phenomena and causal mechanisms,[21] everything from what would happen if someone tried to kick a bowling ball to what it feels like underwater, from how an oven mitt can keep them from getting burned to how a source of heat or light or odor feels stronger up close than far away, and so on and on. It is a rich variety of knowledge and experience, highly fragmented, that students use all the time as they interact with the physical world. Reasoning about questions in physics, students should draw on those resources. In many cases, the ways they are inclined to draw on those resources lead them to wrong conclusions. Students' reasoning that current is used up in light bulbs, for example, draws on resources that would be productive for thinking about how fuel is used up in gas lanterns. But the solution for a student thinking of current being used up is not for them to stop using their common sense. It is for them to find other aspects of common sense to apply, other resources in their repertoire, such as those they would use to understand how trying to hold a moving rope can burn someone's hand. Rather than set their common sense aside, students should search within it for other possible conceptual anchors.[22]

A resource-based model of conceptual knowledge takes a dynamic view of thinking that is in apparent contrast with accounts of novice understanding in terms of coherent "naïve theories" and misconceptions.[23] Research on the latter has established patterns of student reasoning that differ from expert understanding, and these findings have been interpreted to suggest that intuitive knowledge is an impediment to expertise. In some important respects, that interpretation is the opposite of what the original research established,[24] which was that novice "misconceptions" represent sensible, intelligent reasoning well-grounded in experience. Much of the difficulty is that the naïve-theories account views intuitive knowledge as unitary, seeing the misconceptions as the *one way* students have for thinking about the topic. But teachers and researchers who have close contact with students know that students have *many ways* of thinking. Common sense does not have a coherent organization; it is made up of many parts, and the "common sense" answer to a question depends on which parts are activated at a particular instant. A resource-based view provides an account of that variability and of how science can genuinely be a "refinement of everyday thinking."

The core innovations of our reform attend explicitly to student epistemologies, that is, to how they understand knowledge and learning in physics, and we start from a resource-based view in this respect as well. Just as students have a vast collection of resources for thinking about physical phenomena and mechanisms, they also have a vast collection for thinking about *knowledge*, about its various forms and sources, how it can arise and be used in various sorts of activities. Just as they use their collection of conceptual resources for experiencing and making sense of the physical world, they use these *epistemological resources* for experiencing and making sense of knowledge and learning. Depending on the situation — from someone asking them for driving directions to their needing to know a phone number, from their wondering about why a friend is misbehaving to their arguing for a political candidate — they use different epistemological resources for thinking about what knowledge entails, the forms it takes, how it arises, and whether it is valid.

We describe this as a matter of how students *frame* the activity in which they are engaged,[25] that is, how they understand what they are doing. In traditional physics courses students often learn to set their everyday experience aside.[26,27] They frame the task as a matter of receiving and rehearsing information, information that need not make sense. A primary agenda in our courses is to help them frame learning in other ways, tapping productive epistemological resources for thinking about sense-making and argumentation, for understanding physics knowledge as a coherent system of ideas rather than a collection of independent pieces of information. We pursue this in ways we describe below, both explicitly in the instructions and advice we give students and implicitly in the structure and design of assignments, lectures, tutorials and labs.

## TRANSFORMING THE CLASS STRUCTURE WITHIN EXISTING CONSTRAINTS

### The Traditional Teaching Environment

Since the perception of a reform depends on what it is being compared to, we describe briefly the traditional environment for algebra-based physics as it was at the University of Maryland when we began the project in 2000.[28] The traditional algebra-based physics class is taught in two fourteen-week semesters covering the topics of "mechanics, heat, sound, electricity, magnetism, optics, and modern physics."[29] Each half of the class is taught to 400-500 students per semester, divided into 3 lecture sections of 100-200 students. Each lecture section is assigned to a faculty member who is responsible for the content, lectures, assigning reading, and homework. Each lecture section is divided into small group sections of 24 students and each small group section meets for one three-hour period per week run by a graduate teaching assistant (TA). The first hour of the period is typically a problem solving recitation; the last two hours are a laboratory. The students purchase a common text, which is typically the source of all homework problems, and a laboratory manual.

Each professor makes a somewhat independent choice as to what specific content to emphasize, within the constraints of the catalog description. While there is some variation, an attempt is made to keep the first semester fairly common, since a significant fraction of students (sometimes as many as half or more) switch from one lecture section to another after the first term. Homework is handled idiosyncratically. Homework may be assigned from the book or from an on-line homework system (e.g., WebAssign), and may or may not be graded. Laboratories are traditional protocol-based with extensive write-ups and step-by-step guid-ance provided. Students work in pairs and create individual lab reports. Ten laboratories are required each term and makeup periods are provided during two weeks of the term in which students can complete missed labs. A separate faculty member is assigned the responsibility for the laboratories and for training the graduate teaching assistants (TAs) in managing the lab.

The lecture faculty are responsible for creating, grading, and managing the examinations for their own students. There are no common exams. Typically there are two to three midsemester exams and a final. Sometimes exams are multiple choice or short answer, but they often include problems and require calculations. A faculty member's recitation-section TAs are typically recruited to carry out much of the grading. Typically the only interaction between the lecture and the lab part of the class is that the same TAs run the lab and recitation sections.

Traditional lectures consist of presentation of material including demonstrations,[30] derivations, and sample solutions to homework-like problems. There is rarely much interaction with the students during lecture. Attendance during lecture varies from instructor to instructor and may range from 25% to 85% of the registered students. Typical recitations are run by TAs as problem-solving mini-lectures, with the choice of problem sometimes guided by student questions. If the recitation does not contain a required quiz and if the TA has been instructed not to solve the current week's assigned problems, the attendance is typically less than a third of the registered students.

One of us (EFR) taught algebra-based physics in this traditional mode for many years "reasonably successfully," meaning there was good attendance in lecture (typically more than 75%), high ratings in end-of-year evaluations from students (above departmental averages for the class), and some anecdotal successes (individual students reporting relief and delight that the course was not as impossible as they expected).

## The Reformed Teaching Environment

During the five years of the project, the authors were the lecturers of record for a semester of the class 11 times.[31] As a result of our reconsideration of the course goals and on the basis of our resource theory of student learning, we reformed each of the components of the class to be explicit about epistemology. Many research-based reforms exist that help to build students' conceptual knowledge. Many of these are based on a cognitive-conflict[32] or elicit-confront-resolve[33] pedagogical model, in which the students are asked to predict and display their "intuitions." Empirical results are then displayed to show these intuitions are incorrect and they then help them to resolve the conflict. It has been our experience, both as teachers and researchers,[27] that this often has the negative epistemological side effect that students learn to consider their intuitive knowledge and lived experience as irrelevant for physics learning; they learn to set it aside, rather than to draw on and refine it. In order to avoid this, we modified each of these conceptually-oriented reforms.

### The Lecture

In reforming the lectures, we implemented three reforms that increased the epistemological emphasis of the class: explicit epistemological discussions and adaptations of the best practices curricula *Peer Instruction*[34] and *Interactive Lecture Demonstrations*.[35]

### Being Explicit About Epistemology

Throughout the course we made the epistemological framing of the course explicit, including through a vocabulary we introduced early in the semester and integrated into lectures and materials. We designed this vocabulary based on previous work in a small seminar class, described in detail elsewhere.[10] One of use (EFR) created and used a series of icons to use in PowerPoint slides and course materials to help reinforce and remind students of the various epistemological framings.[36] The terms include: *shopping for ideas, sense making, seeking coherence, restricting the scope, choosing foothold ideas, and, playing the implications game.*

*Shopping for ideas* – If the overarching message of the course is that "the whole of science is nothing more than a refinement of everyday thinking," a core activity of the course needs to involve students becoming more familiar and critically aware of everyday thinking. We use the metaphor of "shopping" to help students think of their own knowledge and experience as having a large inventory of possibilities through which they could browse, and we explain it with a story to connect to everyday epistemology:

*Imagine you have met a new person and there's something about him that bothers you, but you can't quite put your finger on what it is. So you think about it, trying to figure out whether he reminds you of someone or you've met him before. You "shop" in your mind, through different sections of your knowledge and experience. You ask "Have I met him before?" and you try out different possibilities: "Have I seen him at the pool? At the store? In art class?" Or "Whom does he remind me of?" again, trying out some possibilities: "Uncle Ralph? Cousin George? Neighbor Charlie?" Eventually you may realize that he looks and sounds a bit like a character in a movie you saw recently. Having figured that out, you know that it's not really this new guy who troubles you but that movie character, and you don't have to worry about it any more. Or, if you were to realize that you've met*

*him before and had an unpleasant interaction, you'd have found that feeling of irritation is warranted.(Ref. 10)*

This sort of shopping in their minds serves two purposes. One is to help students locate the origins of an impression they have about some problem ("I can't explain why, I just think that's what will happen.") as well as to help them locate alternative possibilities ("We're saying electricity *flows*; maybe we should think of other things that flow and try to compare?").

*Sense making* – Students in many college science classes have the view that science is a collection of unrelated facts, and that those facts do not necessarily need to be comprehensible.[26,27] A dramatic example took place in a videotape of a lesson trying to help students build analogies for thinking about electric current. One student asked that the TA stop messing around with analogies and tell them how current <u>really</u> worked. The TA responded, "What do you want me to do, give you a bunch of words that you don't know what they mean?" The student answered (with a straight face), "Well, that's what I'm used to." We emphasize to students that the principles, definitions, and equations of physics should *make sense* – that they should be able to restate that principles, definitions, and equations in their own words and be able to explain clearly what they are saying.

*Seeking coherence / safety net* – Physicists take for granted that knowledge in physics should cohere: We come to accept ideas and findings as true because of the way they hold together with other ideas and findings, ideally with *all* the other ideas and findings of which we are aware. When there are conflicts, we need to resolve them, and if we cannot, it tempers our confidence in the conclusions and our satisfaction with our understanding. Unfortunately, many students are more accustomed to thinking of physics knowledge as a set of facts and formulas, independent pieces of information to remember, and they have come to frame learning science as a matter of memorizing information.[10,16,37] We try to guide them to a more productive framing, again looking for resources in their experience that might help. Students know, for example, the vertical and horizontal lines of crossword puzzles have to be consistent with each other; they understand how game rules need to be internally consistent, as do societal laws.

We also stress to students that their "one-step" recall memory can be unreliable. The mind reconstructs memory from bits and pieces and something remembered may cross up distinct memories.[38] Having coherence means there is numerous cross-linking in memory that provides them with a *safety net* that provides a stability and consistency to their reasoning – a stability that is not present when they memorize independent results.

*Restricting the scope* – One of the challenges for students about learning physics is learning to ignore some of what happens in the real world in order to construct models. If students frame learning physics as learning about the real world all at once, they will constantly be frustrated and confused by the routine practices in physics of making simplifying assumptions, positing idealized conditions and ignoring some aspects of the physical world. We make this an explicit topic of discussion, how cordoning off a portion of the world for attention can serve as a step toward understanding the world more generally, and we make a point of marking when this is taking place.

*Choosing foothold ideas* – Students very often find themselves in the position of not knowing what to believe; that, of course, is common in professional science as well. We introduce the notion of a *foothold idea* as one we choose to accept as true, at least for the time being, as a way to proceed. As we find other ideas and findings fit with a foothold idea, and as we are able to respond to counter-arguments and counter-evidence, we form greater and greater commitment to the foothold; we are willing to work harder to reconcile other reasoning to fit with it. For example, if an experiment produces evidence of fusion taking place at low temperatures, in contradiction to high-commitment foothold ideas about nuclear and atomic physics, or if measurements show that the expansion of the universe is accelerating in contradiction to foothold ideas about the make up of the physical universe, we maintain skepticism of the results and work hard to discredit them, to reconcile the contradiction in favor of the footholds. Sometimes, it becomes too difficult to reconcile the contradictions with current foothold ideas, and scientists choose new ones.

*Playing the implications game*– Having chosen a foothold idea, we consider its implications; if X is true, what would that mean? Often that leads us to something we can't accept, and we abandon X. Sometimes it leads to surprises that turn out to be true. Again, this is a form of reasoning well within students' abilities that they may not apply to learning physics without prompting, if they frame what they are doing as taking in and remembering information. We identify the "implications game" to let students know that's what we're doing

### Peer Instruction (Clickers)

Starting in 2002, we adapted elements of the *Peer Instruction* (PI) environment[35] for this class. Each student was issued a remote answering device (clicker). The instructor periodically asked a multiple-choice question during the lecture to which the students responded using these devices. A computer automatically displayed a histogram of the results.

In the original PI environment, a clicker question typically follows a 10-15 minute lecture segment. The student is asked to think about the answer individually, and then click an answer. If the question is well designed, the class should display a mix of answers. The students are then given two minutes to discuss the question with their neighbors, and the students again click an answer. If there is now large-scale agreement among the students on the correct answer, the lecturer goes on. If not, he or she adds a brief lecture segment to explain the correct answer.

We modify this reform as follows. In all cases, the instructor draws on the class for discussion of the question segment; sometimes the instructor presents the question alone and asks the class to suggest possible multiple-choice answers. In part to help save students from embarrassment, and in part to encourage the habit of mind, we ask students to generate answers and reasoning that they think someone who had not studied physics might believe. Discussion often focuses on students' intuitions based on their real-world experience. After the first click, students have the opportunity to defend or challenge answers (not necessarily their own). Once the correct answer is known, further discussion focuses on the wrong answers, why they were chosen, and whether even they had a "correct" intuitive core. The goal is to encourage students to not just "know" the right answers, but to perceive them as both plausible and intuitive. One of us [DH], often creates spontaneous clicker questions "on the fly" in response to a student question or to a sticky point in the lecture. The other [EFR] tends to have pre-prepared questions and to include them (without answers or discussions) in his PowerPoint slides, handouts of which are distributed via the web the night before the class.[39] For examples, see the supplementary on-line materials, figures (S3) and (S4).

*Interactive Lecture Demonstrations*

We increased the interactivity of our lectures by adapting the *Interactive Lecture Demonstration* (ILD) environment.[36] In the original form of this environment, a full lecture period is devoted to a set of connected demonstrations. The choice of topics relies heavily on education research to determine critical areas where students tend to show or develop misconceptions that interfere with their understanding of the physics being presented. The method relies heavily on cognitive conflict, with an elicit-confront-resolve instructional model. Students receive two identical worksheets, one for their predictions, made after the experiment has been explained but before it is carried out, and one for the results they observe in the experiment. At the end of the demonstration, the students hand in their predictions so as get credit for having participated in the ILD. They keep their results sheet.

We liked the ILD model for its interactivity, its proven success in developing conceptual knowledge, and its guided inquiry structure. We were, however, concerned that the method might work against our epistemological goals. Applied systematically, the cognitive conflict approach can send students the message that their intuitions about the physical world are generally misleading and irrelevant to physics class. This may contribute to what we have documented in students' epistemologies, that they have learned to set their intuitive knowledge aside, rather than to refine it.[26,27] The two-worksheet structure of ILDs embodies that view: Students hand in the page with their intuitions (perhaps to show the instructor how wrong they were before the lesson) and they keep the page with the "right answers". Although we there is no evidence that this kind of activity directly contributes to the kind of problems we have observed in other environments, we wanted our course to send a consistent "meta-message" to students about their intuitions and how to use them in this class.

We therefore modified the approach so that students receive only a single worksheet that emphasizes finding the valid content of a student's intuition and refining it. The lecturer guides students through the worksheet and leads a discussion about the issues it raises. We developed about a half-dozen worksheets to be used in each semester.[40] An example of an "epistemologized" discussion from an ILD worksheet is shown in figure (1). In this example, discussion of the well-known "misconception" that blocking half a lens will result in blocking half the image (instead of reducing the intensity but showing the full image)[41] is paired with discussion of blocking half the bulb – which does result in blocking half the image. By presenting the two situations, we hope to help students refine their existing intuitions, and to avoid implying that their existing intuitions are systematically wrong.

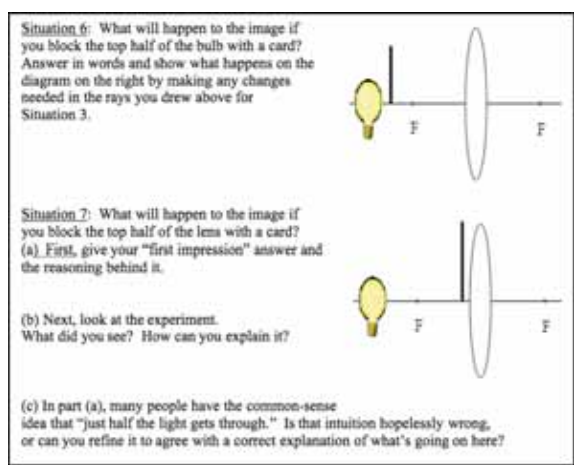

**Fig. 1**: A component of an epistemologized ILD worksheet. The worksheet is done in lecture, students discuss the issues among themselves, and the instructor leads the discussion and shows the demonstrations

The students are not graded on their answers to either the clicker questions or the ILDs, but they are given participation points for doing them. We design homework and test questions to help them assess their understanding of the material discussed during ILDs.

*The Recitation*

In our previous experience with large lecture introductory physics classes over many years, we had observed many TA-led problem-solving recitation sections. In our informal experience sitting in on recitations and interacting with the over many years, we felt that the solutions developed and presented by the TAs seemed to undermined the approaches we were trying to foster in lecture and occasionally contained conceptual or mathematical errors. The students rarely engaged in questioning the TA's presentation in the depth that would have made it useful to them. Current research in our group supports these informal impressions.[42] Indeed, unless we required attendance at recitation through some sort of graded quiz, we found that the number of students attending recitation dropped precipitously – typically to 1/3 of the registered students or fewer. We therefore felt little hesitation in replacing the problem solving recitation by a conceptual tutorial. Other mechanisms were created to provide students a venue to get questions answered about the homework. These are described in the section "Homework".

*Tutorials*

Instead of a TA-led recitation, students work through worksheet-based group-learning activities based on the model developed at the University of Washington.[43] We began with the set of tutorials developed at UW[45] and the Activity-Based Physics Tutorials developed at the University of Maryland.[44] These tutorials were originally designed to produce conceptual gains and have been demonstrated to succeed in this goal,[45] but our previous research indicated that they did not help with the development of better epistemological attitudes associated with the class.[27] We conjectured that part of the problem is the cognitive-conflict approach that stresses the failure of everyday intuition.

So we created some new tutorials, again with epistemologies in mind. Our "epistemologized" tutorials emphasize the reconciliation of everyday, intuitive thinking and experience with formal scientific thinking and to encourage explicit epistemological discussions about the learning process. A common tool that we employed was the paired-question technique developed by Elby.[9] In this approach, instead of offering students introductory questions that research has shown most will get wrong, a pair of matched questions are created so that most students will get one right and one wrong.[46] Elaboration and analysis of the pair of answers shows students that their intuitions are leading them into a conflict. They are then guided to find their (essentially correct) "raw intuition" that underlies both their answers and are guided to maintain that intuition and to refine it in a way that leads to consistent results that are consistent with the physics they are learning. In this way, we hope to convince them that their intuitions about the physical world are valuable and, when properly refined, support the physics knowledge they were learning. An example of a "reconciliation diagram" from a tutorial on Newton's second law (unbalanced force goes with acceleration, not velocity) is shown in figure (2)

Otherwise, the sections proceed in traditional UW Tutorial fashion, with two trained facilitators wandering the room, listening, asking questions, and checking results.

*The Laboratory*

Although our original plan did not call for reforming the laboratory, watching videotapes of students responding in lab changed our minds.[47] We observed students, including the best students in the class, "going through the motions" in following the explicit protocols given in the lab manual. They spent little or no time trying to make sense of what was happening or trying to relate either the procedures or the results to the physics they were learning in the class.[48,49] Students even made comments to the effect that they did not expect to make sense of what was happening, which some students found distressing. In this way, the laboratories sent students messages about the nature of physics knowledge and how one acquired it that contradicted the ones we were trying to send. We therefore spent the period of the project developing and refining the laboratories.[49] The final result was the Scientific Community Labs.[50,51]

*Scientific Community Labs*

Reformed labs are held for periods of two hours with 20-24 students and one TA. The goal is to help students understand the construction of knowledge through measurement and analysis. Instead of an already set-up apparatus and detailed lab manual, we give students a half-page instruction sheet containing a one-sentence question that can be answered empirically, such as "What affects the acceleration of a rolling object?" or "How does the force between two magnets change if you change the distance between them?" The students' task is then to design an experiment using available equipment, make measurements, analyze the results, and present them to the class. We choose questions that make designing and carrying out the experiment feasible.

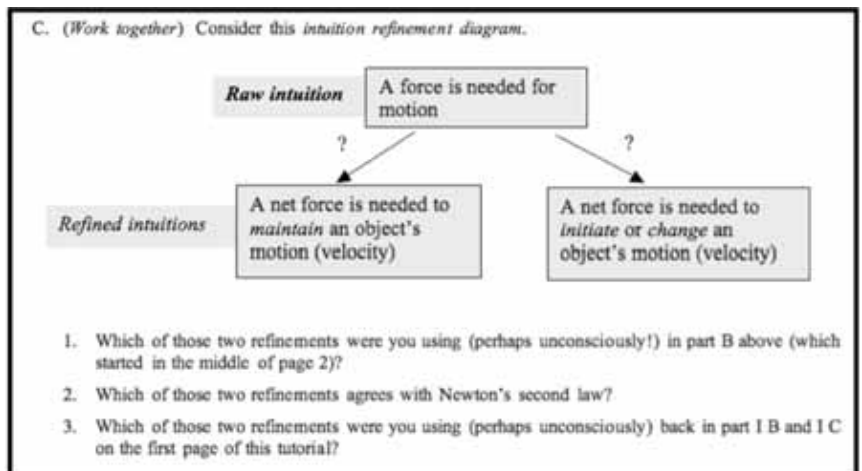

**Fig. 2**: A section from a tutorial worksheet containing a reconciliation diagram and an epistemological discussion. In the actual worksheet, space is left for the students to write their answers.

Unlike traditional practices, we do not use labs as ways to follow up on theoretical treatment in lecture. For example, the lab shown in figure (3)[52] takes place well before rotational motion is discussed in lecture. In this way, we keep the purpose of the lab centered on students understanding of physics as an empirical science. The labs can introduce students to a topic phenomenologically, as preparation for later theoretical development, as is often the order of things in real scientific research and discovery.

Students work in groups of four, write group reports and are evaluated on their thoughtfulness, persuasiveness, and understanding of measurement concepts, as well as on the clarity of their discussion of how they could improve their experiment were they to repeat it. During the lab session, we encourage and there is considerable interaction and discussion among the students and with the TA. Finally, we give students two weeks per laboratory, which gives students four hours to plan, implement, analyze, and discuss each experiment.

There is a practical benefit to for TAs: That students collaborate on reports in groups of four, and that they hand in reports every other meeting, reduce the number of lab reports the TA needs to grade by a factor of eight. Those readings, then, can be quite thorough, and TAs have time to give detailed feedback

*Homework*

Traditional algebra-based physics classes typically assign a large number of end-of-chapter problems for homework. Many are exercises that focus the process of manipulating equations and coming up with numerical answers. Other more substantial problems may also be assigned. Early in the project, we noted informally that many students concentrated their effort on the exercises, which they could do without much thought or understanding, but they would give up on the harder problems before getting very far. That is, we saw our students behaving in ways Schoenfeld observed in mathematics classes, that students seemed to believe they should be able to solve "any assigned problem in five minutes or less."[53]

We decided to drop all exercises, and instead to design homework assignments entirely around challenging problems that require students to think out and make sense of the ideas. Of course, students have experience with problems sets before our class, and many still attempt to use "exercise methods" on more challenging problems,[16] not expecting they would have to spend much time working out problems outside of

**Lab 4: Let It Roll**

**Question:**
*What affects the acceleration of a rolling object?*
   Choose **one** property to investigate as a group.
   Pool your data as a class and try to decide which factors affect the acceleration and which don't.

**Fig. 3**: A typical laboratory handout. The students were also given suggestions for how much time to spend on the various components of the experiment, but no more detailed suggestions as to what to do were given

class for an introductory course. For this reason, we emphasize early and often in the course that they should expect to spend anywhere from 15-60 minute per problem, ideally working in groups, discussing the issues with each other.

In accordance with this expectation, we only assign about five problems each week. They include a mix of challenging activities including representation translation problems, context-based reasoning problems, ranking tasks, estimation problems, and essay questions with epistemological content.[54] Examples of the first two types of questions are shown in the supplementary on line materials, part 2, figures (S5)-(S6). A collection that includes many of the problems we have developed and used is available on the web.[55]

We have had to compromise substantially on the extent and quality of the feedback we can give students to their homework. In order not to overload the TAs, we pick one problem each week for careful grading on a five-point range with written feedback. TAs grade other problems "lightly," without written feedback, for what we describe as "honest effort." We provide elaborate solutions on the course website after homework is due and enjoin students to study them; students need to learn that a good score on their assignment did not necessarily mean that they had done the problem correctly due to the "light" grading.[56]

*The Course Center*

Since we converted the traditional discussion sections to tutorials, they no longer provide opportunities for students to discuss the homework problems. To close this gap, we set up a "course center," staffed by TAs or the instructor approximately twenty hours per week, where students could gather to work on homework. The TA or instructor offers assistance and coaching on good problem solving strategies, not solutions. After trials in a number of different classrooms in different semesters, we found that it was necessary to arrange the furniture in the room to discourage the TAs from making presentations to the entire class. There is now no central writing space, and the tables are arranged so that the students sit facing each other. White boards are available, but only at places behind the tables where seated students could easily reach them but the TAs cannot.[57]

*Exams and Quizzes*

We design exams to require the kind of thinking we want students to learn, and we deliver them in a way to communicate that they are to be used as formative rather than as purely summative evaluations. The exams and quizzes must include items that call on the students to use the epistemological skills we are trying to help them develop.

One of us (EFR) follows a strict structural pattern on hour exams and explains it to the students. Every exam contains 5 questions: A multiple-choice multiple-representation question (worth ~25 points), two long-answer problems (worth ~25 points each), an estimation problem (worth 15 points), and an essay question (worth 10 points). On the long-answer problems the answers are worth only 5-10 points while the explanations and reasoning are worth 15-20 points. On the estimation problems, the answers are worth only 3 points (and a wide range of answers were acceptable) while the method is worth ~10-12 points. The estimation questions require the creation of numbers from one's personal experience and the experience has to be explained in order to get full credit. The reduced emphasis on getting an accurate answer and the added stress on reasoning is in response to the tendency of students in this population to focus <u>only</u> on answers and to ignore reasoning. A sample multiple-choice multiple-representation question, essay question, and estimation problem are shown the supplementary on-line materials, part (2), figures (S3), (S5), and (S6).

Exams were typically given in the last class of a week, graded over the weekend, and returned to the students in the first class of the next week. The exam with the grading pattern was then gone over in detail during that class. The discussion included not only the "right" answers but also a discussion of the common errors, misconceptions, and difficulties that many students encountered.

*Makeup Exams and Regrading* – One challenge is that most of the students in the class are juniors and seniors with considerable experience in other science classes, with expectations about what we would (or even could) put on a test. Talking with students at the beginning of the project, we often heard something like, "Well, science exams have so much time pressure, that you really have no time to think during an exam and the profs can't expect you to. So, you have to memorize stuff so you can give it back quickly on an exam."

These expectations, we believe, help explain why many students do poorly on the first exam in the course. For them, the feedback is negative and distressing, about 1/3 of the way into the semester, with some critical material behind them and most of the semester left to go. One option would have been to give three hour exams a term and drop the lowest grade, but this could send the message that if they did poorly on the first exam they could let that material go. To the contrary, physics learning is highly cumulative; we need them to think it is important to go back and learn the material if they have not understood it well.

An important goal of the exam structure is to help the students learn to use their exam results to focus on identifying problems in their thinking and in their approach to learning. We want to encourage them to look at the problems they missed and ask not only, "What is the right answer?" but "Why didn't I get the right answer?" We do this with makeup exams and regrading.

The makeup is given outside of class at the end of the week following the original exam, and if students are dissatisfied with their grades on the main exam they may take it. If they do, then they receive the average of the two grades on the two exams. This means that they could actually lose points by choosing to take the makeup. We attempt to make the makeup exam as close in difficulty to the original as we can, but do not give the same problems.

We also explain to students that in our experience, those who simply study again as they had studied for the first exam had an equal chance of going up or going down. Students who study by focusing on <u>why</u> they had missed problems and refining their thinking and understanding almost always improve, sometimes very substantially.[58] Typically, 25% of the class chooses to take the makeup with 80-90% of those improving their grade significantly.

A second technique we used to focus students to think about their own thinking is the regrade. When we go over the exams in class we stress that the graders had many papers to read and might not have understood a particular student's reasoning. Students who believe that they should have had a higher score on any problem may write a page with an explanation of their reasoning and argument for more points. Papers that are handed in with statements like "Please look at problem 3 again" or "I think I should have had more points on problem 2" are returned with instructions to write an explanation discussing their answer and the correct answer. We intend this to focus students on their own thinking and how it compares to the solution shown in class. Having to write a detailed explanation, we hope, can help students find where they went astray, as well as provide the instructor with an opportunity to interact one-on-one with students who have specific difficulties. Typically, about 30% of the students write requests for regrades.

*Quizzes* – As an attempt to give students still earlier opportunities to change their expectations about the course, we decided to introduce weekly quizzes starting from the second week of class. These quizzes are given during the first 10 minutes of the first class of the week and typically focus on applying the processes learned in the previous week's tutorial to a new example. (An example is given in the supplementary on-line materials, part 2, fig (7).) This has the advantage of both focusing students on the value of the tutorial early[59] and on demonstrating that memorizing answers is not effective in this class.

The quizzes are collected and the answers given (without explanation). After class, the quizzes are graded, the specific answers given by each student recorded, and the quizzes handed back in the next class. That class began with a discussion of the quiz and a presentation of the distribution of answers chosen by the class. Students are asked to justify common answers and to discuss reasons for choosing them and ways for evaluating their own thinking to understand how to know they were wrong or right.

## Coherence and Synergy

We made every effort to try to get all the parts of the class to work together and send the same epistemological message. Tutorials and ILDs often included the same epistemological icons used in lecture. Examples in lecture often referred to the tutorials. And the lecturer often spoke about issues related to the laboratory or homework.

## DATA COLLECTION AND OBSERVATIONS

We collect data about student understanding and epistemologies in a variety of ways, for purposes of both research and instruction. The data includes video of students working in tutorials, laboratories, and the course center; records and scans of student responses on clicker questions, quizzes, homework and exams; pre and post class interviews with individual students, and pre and post class surveys, both of conceptions and epistemologies.

*Video* – The project produced approximately 400 hours of videotape of students participating in tutorials, approximately 500 hours of videotape of students participating in laboratories, and approximately 50 hours of students working in Course Center to solve homework problems.

*Artifacts* – Weekly homework, lab reports, and exams were scanned for approximately 500 students throughout the four years of the project.

*Interviews* – Approximately 30 one-hour semi-structured interviews were collected from volunteers at the beginning and end of the first and second terms in the first two years of the project. These were used to help us understand students' initial expectations and to see how well our reforms were achieving our goals. As a result of observations of the video data of students working in tutorials, a student with particularly interesting epistemological orientation was sought out and volunteered to provide 6 hours of interviews as a case study.[14] In another instance, we interviewed a student whose performance improved dramatically between a first-hour exam and the make-up, to understand his sense of the reason for the improvement.[7]

*Surveys* – All students taking the project class took surveys at the beginning and end of the first term and at the beginning of the second term. Surveys included a mechanics conceptual survey (either the FCI or FMCE) and an attitudes/expectations survey (a combi-nation of the Maryland Physics Expectations (MPEX) Survey and the Epistemological Beliefs Assessment for Physics Science (EBAPS) Survey). These are discussed in more detail in the section on results.

## RESULTS

In this paper, we give an overview of the evidence of progress students made with respect to our goals. We also discuss some interesting and relevant instances. We discuss these in four categories: concept learning, laboratory behavior, attitudes and expectations, and student perceptions of the class

## Concept Learning

The primary goal of most reformed classes in first semester introductory physics is good conceptual learning of mechanics.[60] In our project, we placed priority on epistemological development. Although we were building our reforms on best-practices curricular materials designed and demonstrated to improve the actual learning of concepts, we were uncertain whether our shifted emphasis would be at the cost of the conceptual gains produced by the materials in their original form. Fortunately, that turned out not to be the case.

*It is possible to obtain strong conceptual gains in a class whose primary focus is epistemological learning.*

We used two widely accepted instruments: the Force Concept Inventory (FCI)[61] and the Force Motion Conceptual Evaluation (FMCE).[62] Although these tests are narrow in scope and test student performance in only a single environment, studies have shown that these are reasonably good indicators of broader student understanding and skill development.[47]
The figure of merit for pre-post testing often used is the average fractional gain, $\langle g \rangle$. This is defined as the fraction of the number of points the class gained compared to the fraction of the number of the percentage points the class could have gained – a kind of educational efficiency.[63]

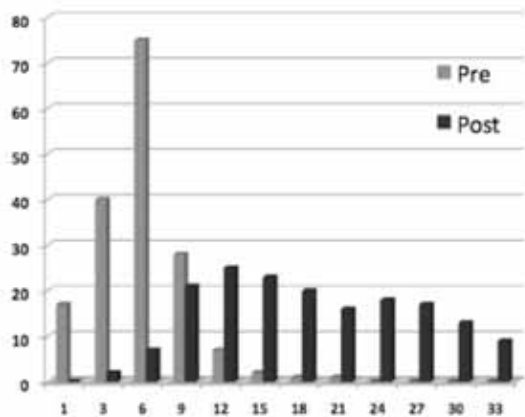

**Fig. 4**: Pre-post distributions of scores on FMCE from the fall 2007 class. (Scoring using spreadsheet by M. Wittmann.[73]) The pre-post averages were 15% and 53%, and the fractional gain was $\langle g \rangle$ = 0.44.

$$\langle g \rangle = \frac{(\text{post \% av.}) - (\text{pre \% av.})}{100 - (\text{pre \% av.})}$$

Typical traditional classes, whether in high school or college, score gains on the FCI of $\langle g \rangle \sim 0.2$ while reformed, active engagement classes tend to score higher – on the order of 0.35 for modest reforms and on the order of 0.6 for more extensive reforms.[47,64,]

During the project, we taught the first semester course four times. A sample of our results are shown in figure 4). The average gains in the class ranged from 0.44-0.47. These results are well into the range of gains shown by the stronger of the "active engagement" classes in the Hake survey (ref. 70) and are the best we have obtained in a large class at the University of Maryland.[65]

We conclude that it is possible to attain strong conceptual gains, even in a class whose focus is on epistemological learning.

*It is possible to help students develop their intuitions that the physics they are learning makes sense.*

Conventional applications of conceptual surveys do not distinguish between two outcomes: (1) Students coming to recognize what the course considers to be the correct answers and (2) students coming to see those answers as making sense to them. Consistent with our epistemological agenda, we hoped to achieve the latter, and specifically to avoid the situation of students learning to provide answers they do not personally believe.

We applied the split-survey task developed by McCaskey, Elby, and Dancy.[66] This task asks students first to "circle the answer that makes the most intuitive sense" and second to "put a square around the answer that [they] think a scientist would give." A typical example of a student response from the FCI is shown in figure 5).

In this example the student shows that she knows the "correct" answer but by splitting indicates that she does not find that answer intuitive. She has not reconciled Newton's Third Law with her sense that the bigger (or more active) object must exert the greater force.[67]

In most semesters, only the first author delivered a reformed class, along side two traditional sections taught at different times. Between the first and second semesters, a significant number of students would transfer from one section to another, typically driven by schedule constraints.[68] In the spring semester of 2003 we gave the split task FCI to the ~200 students entering the second semester class. Approximately two-thirds of the students had taken the reformed first semester class while the remaining one-third had taken a traditional one.

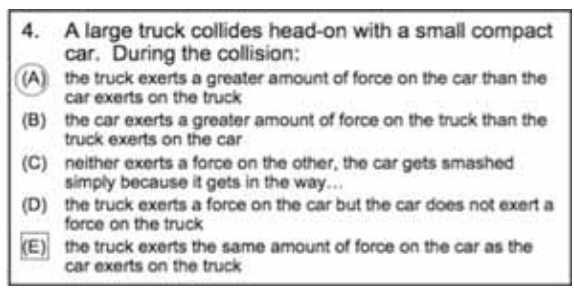

**Fig. 5**: A "split" response of a student on an FCI

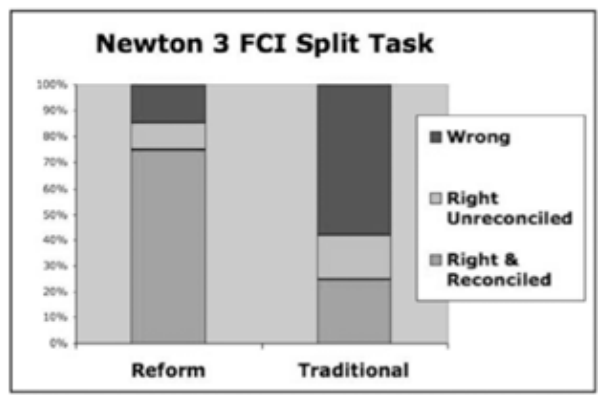

**Fig. 6**: Split task results on the N3 cluster of the FCI. "Right and unreconciled" means the correct answer was given as the scientist's answer but a different answer was given as the intuitive one.

The results on the Newton's-third-law cluster of questions on the FCI (four items) for the two subsets of students are shown in figure 6). We see that not only did the students from the reformed mechanics class answer a larger percentage of the questions correctly (~85% compared to ~45%) but the correct answers were split less often (~90% compared to ~60%).

These results are dramatic with the number of "right and reconciled" answers increasing from ~25% with the traditional approach to ~75% with our reformed approach. Although there are many issues to be dealt with in order to establish a convincing research result,[69] the results of these preliminary data suggest that something important is happening in the reformed class that is not happening in the traditional one.

### Activities in laboratory

*Students tended to frame our reformed laboratories as about sense-making, in contrast to simply following instructions, as students tend to frame traditional laboratories.*

In the spring of 2001 we videotaped students working in traditional laboratories (working from detailed instructions in a lab manual prepared by Physics Department faculty).[70] We analyzed these videotapes with respect to what students treated as the source of knowledge (see refs. 49 and 51for evidence and coding). The analysis showed that students in traditional laboratories access knowledge almost solely from authority (the manual, the textbook, the TA). There is almost no construction of knowledge, coherence building, or sense making. Data taken from the equipment was treated in an authoritative way, as "the right answer (unless we made a mistake).[49]

To try to overcome the barrenness of student epistemological response in this environment, early in the project some TAs were given the flexibility to vary the lab structure. When the TAs chose to take the lab manual away and give students only the task and the equipment, the students seemed to respond by accessing much richer sources of knowledge including social (group) construction and sense making. We decided to undertake modifications of the laboratory not originally planned as part of the project and created the Scientific Community Labs (SCL) described above. These laboratories present the students with a task and equipment but no protocol. Students have to build an understanding of how a measurement can answer a question. The lab's social interactions are explicitly structured and the lab tasks are designed to help foster understanding of the nature of measurement.

In order to determine whether these new laboratories were meeting our epistemological goals, we analyzed student discourse, coding student statements by four categories:[51]

- *Sense making* – statements that connected the lab to students' sense of physical mechanism.
- *Logistics* – statements that concerned how to proceed without specific reference to ideas about the physics.
- *Off-task* – statements not relevant to lab activities or to the related physics.
- *Metacognitive* – statements expressing a feeling or an evaluation about some thinking someone had stated.

Statements in the first three categories were typically part of an extended discussion lasting minutes or more. Metacognitive statements were typically a single statement. ("Wait, that doesn't seem right." "I don't get it." "But doesn't that contradict what he said in lecture?") We therefore display the results as an adaptation of the time-block plot of the kind used by Alan Schoenfeld.[71] The videotape is transcribed and each student statement is classified into one of the four categories. Times are then blocked on the figure for the first three categories. When any student makes a metacognitive statement, it is marked as a small triangle attached to the bar associated with the next statement. If the metacognitive statement leads to a change to sense-making mode, it is circled. (For more discussion of the protocol used in constructing these plots, including coding and inter-rater reliability.[49,55,72])

This analysis shows the reformed, science community labs lead to dramatic increase in the time students spend sense-making (see fig. 7). The fraction of time spent in sense-making in the traditional lab was very small – typically 5% or less – a total of 5 minutes or less in a two-hour lab. Most of the attempts at sense making in those labs were brief and unsuccessful. In the SCL, sense-making increased to about 20% – 20-25 minutes in a two-hour lab. Often, the sense-making lasted for a number of minutes.

Even more interesting is the change role of metacognitive (evaluative) statements. They occurred about equally often in the two labs, but in the traditional lab they were rarely productive: Negative evaluations of sense ("I just don't get this.") were rarely succeeded by sense-making; students would simply continue to do what they had been doing. In the SCLs, metacognitive statements often led to a period of sense-making.[79]

## Attitudes and expectations

The previous results show that in the reformed classroom, students learned concepts, sensed the coherence of the physics they were learning with their intuitions, and spent more time in their laboratories seeking cogency. These are all measures of how students are functioning in their learning. We were seeking development along these axes as well as in students' awareness of epistemological issues.

One way to measure students' epistemological progress is by survey. Surveys of this type include the Maryland Physics Expectations (MPEX) survey,[27] the Epistemological Beliefs Assessment for Physical Sciences (EBAPS),[73] and the Colorado Learning Attitudes About Science Survey (C-LASS).[74] The MPEX and C-LASS surveys consist of a list of statements with which the students are asked to agree or disagree on a 1-5 point scale, strongly disagree, disagree, neutral, agree, and strongly agree. The EBAPS also contains such items but adds a set of *scenario* items. In these, students are presented with two scenarios and asked to decide which would be more effective in helping them learn physics.

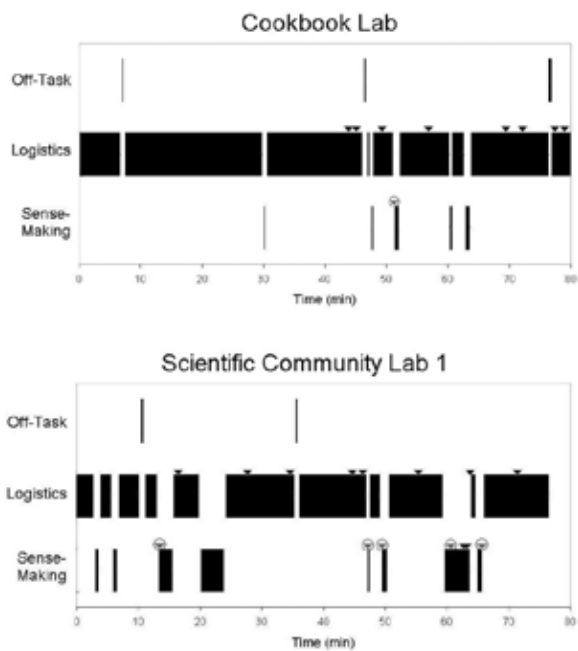

**Fig. 7:** Discourse analysis of a traditional "cookbook" lab and a scientific community lab showing a track of the time when group discourse was off-task, engaged in logistics, or focused on sense-making. Metacognitive statements are marked by small triangles. Metacognitive statements that are productive in changing the mode are circled. These results are typical. (From ref. 49.)

The items are clustered into a variety of categories, including *concepts*, *coherence*, *reality*, *mathematics*, and *independence*. On some items, experts would prefer that the students agree with the item, on others that they would disagree. If the student's response agrees with those preferred by an expert it is said to be *favorable*, if it disagrees it is said to be *unfavorable*.

These three surveys have been tested in large lecture classes with thousands of students. The results are highly consistent. Students typically enter with attitudes that agree with those preferred by experts at a level of about 50% to 65%. After one semester of introductory physics, these attitudes deteriorate by 5-10% or more, whether or not the class has been reformed to produce improved conceptual learning. The reality link is a particular problem, dropping by 10-20%.[27] It has, however, been demonstrated that in a small class with a strong emphasis on epistemology, substantial gains can be obtained on such surveys.[9]

Since different populations require somewhat different surveys, we created a new survey for this study that included elements from the MPEX and the EBAPS. This survey, which we refer to as MPEX-II, is included in the supplementary on-line materials (part 3) with the assignment of the elements to categories. A paper on the construction, validation, and detailed results of the survey is in preparation.[75] Four sample items, their source, their category, and their polarization (whether an "agree" response is favorable or unfavorable) are given in table 1. The main result is dramatic.

**TABLE 1.** Some examples of items from the attitude / expectations survey.

| Item | Source | Category | Favorable Polarization |
|---|---|---|---|
| When solving problems, the key thing is knowing the methods for addressing each particular type of question. Understanding the "big ideas" might be helpful for specially-written essay questions, but not for regular physics problems. | EBAPS | Concept | Disagree |
| A significant problem in this course will be being able to memorize all the information I need to know | MPEX | Coherence | Disagree |
| When learning a new physics topic it's important to think about my personal experiences or ideas and relate them to the topic being analyzed. | MPEX | Coherence/ Reality | Agree |
| Let's say a student has limited time to study, and therefore must choose between the following options. Assuming the exam will be a fair test of understanding, and assuming time pressure during the exam isn't an issue, which option should the student choose?<br><br>(a) Learning only a few basic formulas, but going into depth with them.<br>(b) Learning all the formulas from the relevant chapters, but not going into as much depth.<br>(c) Compromising between (a) and (b), but leaning more towards (a).<br>(d) Compromising between (a) and (b), but leaning more towards (b).<br>(e) Compromising between (a) and (b), midway between those two extremes. | EBAPS | Coherence/ Math | (a)-(c) |

*It is possible to achieve significant gains on an Expectations/Attitude survey in a large lecture class without sacrificing conceptual gains.*

The pre-post matched survey results on the MPEX-II (N=146) are shown in figure 8). The class shows strong gains in the categories of concepts and coherence and essentially no change (a non-significant gain) in independence. Note that the difficult-to-improve sub-category "reality" improves from 66% favorable to 73% favorable. The favorable percentages in each of the items in table 1 each improved by 30% or more.

These results suggest that not only did the students improve on the functional aspects of their epistemologies in the class, they were aware of and could recognize these changes.

## Student perceptions of the class

In addition to the pre-post surveys we carried out, we have additional information about the perceptions of our reforms by students: a survey carried out by the university administration that happened to fall during one of our reform classes and some particular instances that illustrate phenomena we observed with a larger number of students.

*A serendipitous external evaluation: The CORE survey*

The University of Maryland has a series of distribution requirements known as CORE. Courses approved for CORE are intended to help students gain "a strong and broadly based education," to "introduce the great ideas and controversies in human thought and experience," and to provide "a strong foundation for…life-long learning."[76] Algebra-based physics has been approved for decades as suitable for meeting the science component of CORE. Every few years, the university's central administration carries out an evaluation of CORE classes, one of which took place during this project.

The evaluation consisted of a survey with eight items answered on a scale of (1) to (5) ("not at all" to "a great deal"). The items were of the form: "To what extent has this course been intellectually stimulating?" and "To what extent have the writing assignments and/or examinations in this course given you opportunities to think carefully and critically?" (The full survey is given in the supplementary on-line materials, part 4.) In addition to doing the survey in our reformed lecture section, another professor who was teaching using the traditional method in another lecture section at the same time also carried out the survey.

The traditional section scored sufficiently strongly to retain our CORE approval (results between 2.5 and 4.0 on the various items). These results agreed almost perfectly (to within 0.2) item-by-item with the survey done in the class five years earlier with a different traditional instructor. The results in our reformed class were consistently almost a full point higher than in the traditional section for each item. These results are encouraging, especially given that (1) the survey was developed independent from the project, and (2) we were unaware that the survey was to be done and therefore did not intentionally orient the class towards the issues being measured.

*Resistance: It can take a while for students to get used to the approach*

Many of the students in this class were juniors and seniors and a significant fraction (about one-third) were pre-meds. What this meant was that many of the students had considerable previous experience in science classes and many had outstanding academic records. They thought they knew what was demanded of them in a science class and they thought they knew how to cope with those demands. When we told them at the beginning of the term that we were "changing the rules," some got nervous, and some got angry.

In one case, one of us (EFR) heard a report from a TA that one student, "Karen" (pseudonym), was being especially difficult in tutorial and course center. Indeed, a few days later he was called into the Chair's office to respond to a parent's complaint about the class. After reassuring the Chair that the class was under control, he passed a message to Karen through the TA to come and see him. When she arrived, her body language suggested a mix of emotions – nervousness, anger, and uncertainty. The instructor listened to the student's complaint calmly and was reassuring. He determined that she was a 4.0 pre-med who felt confident of being able to succeed in a traditional class and was unsure of being able to meet our new expectations. He reassured her that we understood what students had to do to succeed in our framework, that we had prepared many instructional resources to help them, and that his door was always open to any student who needed help.

After a somewhat slow start on homework, Karen achieved an A on the first exam and continued to perform strongly, earning an A both semesters. At the end of the year, EFR received a letter from her with the following comment.

*Your class was one of the most interesting and beneficial classes I've taken at the University – I improved my thinking skills, creativity, and teamwork skills, not to mention that I learned a lot of Physics! Your style of teaching was one that I feel lucky to have benefited from…and my mom thought so too! … So I'd like to first express my gratitude to you for providing that experience.*

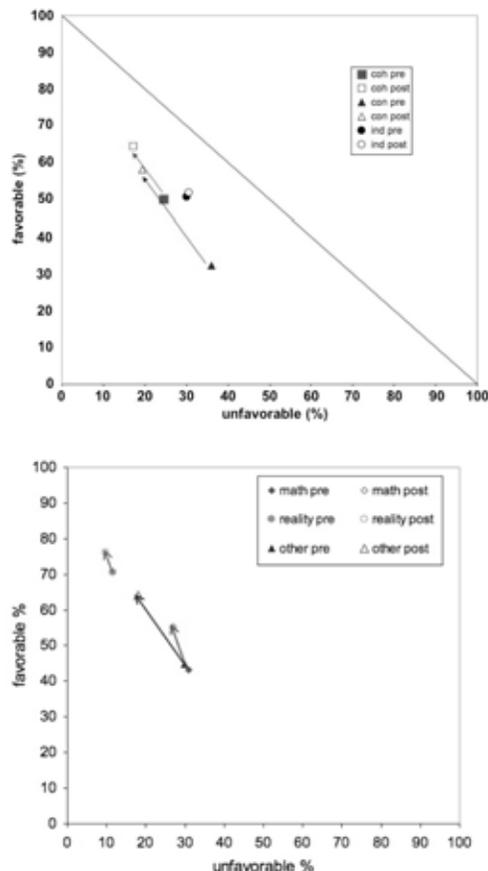

**Fig. 8**: The results on the MPEX-II in the first semester of one of our reformed classes. Other semesters (and our other instructor) looked similar. a) The main categories of concepts, coherence, and independence. b) Subcategories of concepts: math, reality, and other.

*Some individuals undergo significant changes in their approach to science*

One extreme case in our other instructor's class (DH) was a student, "Louis," who failed the first midterm exam with a score of 36%. This prompted him to meet with the professor and ask what he was doing wrong. A conversation took place in which the instructor advised the student to try to make sense of the material by considering "how he would explain it to a ten-year old." A week and a half later Louis took the make-up exam and his score jumped to 84%, the highest on the make-up and near the top of the original distribution. In a videotaped interview with a researcher not involved in teaching the course he explained that his interaction with the professor prompted him to change his approach to studying, from "memorizing the book [and] every word of those homework solutions" to trying to "write down an explanation like to a ten-year-old," using analogies to everyday ideas.

Six months after the course was over, Louis initiated a correspondence to tell the professor "that since I've taken your class, I have a 4.0 GPA, compared to a much lower GPA before your class. I think this increase in GPA has a lot to do with the things I learned in your class — not about physics, but about learning in general." In the exchange that followed, Louis explained why the advice to "explain it to a ten-year-old" had been effective for him: He had experience tutoring, both children and peers, and at the time he was enrolled in the course he was working as a tutor, using strategies of trying to connect to what his tutees already knew. The professor's advice made him more aware of what he was doing as a learner and connected to epistemological resources he possessed, but was not previously making use of in his physics class.[7]

## DISCUSSION AND CONCLUSIONS

In this paper we have summarized a series of reforms for algebra-based physics intended to explicate the epistemological elements of the implicit curriculum and to provide students explicit instruction in learning how to think about science and to understand the process of scientific reasoning. The transformations of the instructional environment were built on existing best-practices curricula that had been demonstrated to provide strong conceptual learning (as measured by conceptual surveys) but to have little effect on expectations and attitudes (as measured by attitude surveys). We reformed these elements based on a theory of thinking and learning, the resources framework. This framework focuses attention on the resources, both conceptual and epistemological, that students bring to class.

## Discussion

Students responded well to these curricular transformations, demonstrating both strong conceptual learning and increased ability both to use and articulate the need for concepts, coherence, and cogency. However, as with any major reform, especially one where student expectations are not met, it is not sufficient to simply introduce the reforms by using the transformed materials. Considerable effort is needed to help students make sense of and become comfortable with what is going on. Our experience suggests three important ways to do this: attend to student framing, be consistent, and restrict the content appropriately

### *Attend to student framing*

Students frame the way they are going to think about our class based on their previous experience in other science classes. If these have been teaching an inappropriate implicit epistemological curriculum, students may not bother to pay attention to statements that this course is going to be different.

Despite explicit statements on the first day of class, detailed handouts explaining the goals of the course, and repeated statements in class, many of our students seem primed to ignore anything we say that they do not interpret as direct content.[77] Despite a statement in the main handout to the effect that homework was the most important learning activity and repeated mentions of it in the class, one student commented on the anonymous faculty evaluation at the end of term, "You won't believe this, but I actually learned the most in this class doing homework!" Despite explicit statements in both the main handout and in class that homework is not graded for feedback and you have to read the posted solutions, some students reported on Pick-a-Prof, that "TA's didn't grade for correct answers on the HW's so you went through the semester THINKING you had the right idea, till you got your exam back and you were wrong."

For an epistemologically oriented class it appears necessary to be continually working to help the students reframe their understanding of what success in the class entails – articulating the class expectations and epistemological goals and listening to (and demanding) their feedback on what is happening.

### *Be consistent*

Traditional courses send unspoken messages about the implicit epistemological curriculum; our reformed course sends very different and (hopefully) explicit ones. The one of us who had been teaching large lecture classes in a traditional mode for nearly twenty-five years (EFR) found himself often slipping "off message" unintentionally. Furthermore, when one of us (DH) taught a semester of the class with traditional rather than reformed laboratories, he found smaller MPEX-II gains than when he taught with a reformed lab.

### *Restrict the content appropriately*

One challenge that faculty considering a reformed class often make to the reformers is, "What did you have to leave out?" The idea that one has to "cover" a particular set of material, whether or not any of the students understand it seems peculiar, but it is widespread. Perhaps it is a way of transferring the responsibility for the success of the class from the teacher to the student. ("I covered it all. It's not my fault they didn't get it.") One might hope that the students who "didn't get it" might at least have good notes that they could look over in later years when they might need the material, and not just "a bad taste in their mouths". Unfortunately, few of our students take good notes and fewer will look at their notes in later years.

An approach that is more appropriate to our goals is to "uncover a little rather than cover a lot."[78] Having large blocks of material in a class that the students cannot expect to understand means lots of material that they will have to memorize. In and of itself, that might not be a problem, but because students have intuitive epistemologies the effect may not be restricted to that particular material: Students may learn or have reinforced the idea that physics does not make sense and apply that to material they could otherwise have understood. We attempted to remove topics that appeared to lead even our top students into confusion despite our best attempts with whatever best practice materials we could bring to bear. We determined this by having final exam questions that included enough open ended and explanatory components to give us some insight into student thinking on these topics.

This lead us to eliminate such topics as heat engines, magnetic induction (Faraday's law), Gauss's law, the details of electromagnetic waves, and much of modern physics.[79] Some of this we did with great regret, as we felt it was important for the population of biologists to know about these topics. However, we accepted the idea that we could not teach everything and that it was epistemologically more effective to have students learn topics in physics that they could genuinely understand rather than be exposed to topics that were interesting but that they could not master in the limited time allotted.

Note that these decisions are highly dependent on instructional tools and environments. It is entirely possible that within the next few years some researchers will figure out how to teach Faraday's law in a way that this population can make sense of with only a few hours of lessons. When this occurs, it will be entirely appropriate to rethink our selection of content.

## Conclusion

In any class, we teach an implicit epistemological curriculum. For many students, what they learn about these issues of the nature of scientific knowledge and what it means to learn and understand science may be the most important things they take away from our class.[80] When these lessons are tacit, inexplicit, and unevaluated, students may learn the opposite of what we would intend. Our reforms in algebra-based physics give an illustration of how an implicit epistemological curriculum can be analyzed, explicated, and evaluated for a particular population of students. A critical issue is an understanding of how students' intuitive epistemologies play a role in their learning. Attending to that issue may prove of value in other classes as well.

## ACKNOWLEDGMENTS

We are grateful to the members of the University of Maryland Physics Education Research Group who were responsible for carrying out the research and development reported here. In particular, faculty Andy Elby, Laura Lising, and Rachel Scherr and graduate students Paul Gresser, Ray Hodges, Rebecca Lippmann, Tim McCaskey, Rosemary Russ, and Jonathan Tuminaro. This material is based upon work supported by the US National Science Foundation under Awards No. REC-00-87519 and REC-04-40113. Any opinions, findings, and conclusions or recommendations expressed in this publication are those of the author(s) and do not necessarily reflect the views of the National Science Foundation.

*Redish & Hammer*  *An Epistemological Curriculum*
[37] D. Hammer, "Two approaches to learning physics," *The Physics Teacher* **27**, 664-671 (1989).

[38] D. Schachter, *Seven Sins of Memory: How the mind forgets and remembers* (Houghton-Mifflin, 2001).

[39] A collection containing some of our clicker problems is available on our website at http://www.physics.umd.edu/perg/role/PIProbs/ProbSubjs.htm.

[40] Available online at http://www.physics.umd.edu/perg/ILD.htm.

[41] F. M. Goldberg and L. C. McDermott, "An investigation of student understanding of the real image formed by a converging lens or concave mirror," *Am. J. Phys.* **55**, 108-119 (1987).

[42] R.M. Goertzen, R. Scherr, & A. Elby (2008), to be published.

[43] L. McDermott, P. Shaffer, and the University of Washington Physics Education Group *Tutorials in Introductory Physics*. (Prentice-Hall, Inc., 2002).

[44] M. Wittmann, R. Steinberg, and E. Redish, *Activity-based Tutorials*, (John Wiley and Sons, Inc., 2004).

[45] E. F. Redish, J. M. Saul, and R. N. Steinberg, "On the effectiveness of active-engagement microcomputer-based laboratories," *Am. J. Phys.*, **65**, 45-54 (1997).

[46] The creation of such a pair requires a good knowledge of both the research database (to identify common misconceptions) and students productive resources (to generate contexts that will facilitate students in spontaneously proposing correct answers.

[47] R. Lippmann, "Students' understanding of measurement and uncertainty in the physics laboratory: Social construction, underlying concepts, and quantitative analysis," PhD dissertation, University of Maryland (2003) [available at http://www.physics.umd.edu/perg/dissertations/Lippmann]

[48] F. Reif & M. St John, "Teaching physicists' thinking skills in the laboratory," *American Journal of Physics*, **47**, 950- (1979).

[49] R. Lippmann Kung, "Metacognition in the Physics Student Laboratory: Is increased metacognition necessarily better?" submitted for publication (2007);

[50] R. Lippmann Kung, "Teaching the concepts of measurement: An example of a concept-based laboratory course," *Am. J. Phys.* **73**:8, 771-777 (2005).

[51] P. Gresser, "A Study Of Social Interaction And Teamwork In Reformed Physics Laboratories," PhD dissertation, University of Maryland (2006) [available at http://www.physics.umd.edu/perg/dissertations/Gresser]

[52] The rest of the page of the handout consisted of suggestions for time management of the lab period.

[53] A. H. Schonfeld, *Mathematical Problem Solving* (Academic Press, 1985).

[54] E. F. Redish, *Teaching Physics with the Physics Suite*, chapter 4 (John Wiley & Sons, Inc., 2003).

[55] The "Thinking Problems" collection is available at http://www.physics.umd.edu/perg/abp/TPProbs/ProbSubjs.htm. Many of these problems have fully written out solutions that are password protected. The password is available to legitimate instructors upon request to redish@umd.edu.

[56] Unfortunately, we have no data on how many students acted on this admonition.

[57] For details, see ref. 16.

[58] We only have informal experience to this effect; we have not conducted a systematic study.

[59] The traditional tutorial model (ref. 43) relies on an ungraded cognitive-conflict pretest and on a tutorial question in each mid-semester exam to achieve similar goals. Our post-test graded model was a better fit to our epistemological goals.

[60] See, for example, ref. 54, chaps. 7-9.

[61] D. Hestenes, M. Wells and G. Swackhamer, "Force Concept Inventory," *Phys. Teach*. **30** (1992) 141-158.

[62] R.K. Thornton and D.R. Sokoloff, "Assessing student learning of Newton's laws: The Force and Motion Conceptual Evaluation," *Am. J. Phys.* **66**(4), 228-351 (1998).

[63] This measure is typically used to permit the comparison of classes with different pre-test scores. Some care must be taken as the score is particular to the specific test used and may distort the result when either the pre- or post-test averages are high, producing end effects. We are grateful to Robert Mislevy (private communication) for this comment.

[64] R. Hake, "Interactive-engagement versus traditional methods: A six-thousand-student survey of mechanics test data for introductory physics courses," *Am. J. Phys.* **66** (1998) 64-74.

[65] Although the other classes we have tested at Maryland have all been calculus-based, the Hake survey shows that for the FCI in the mid range (pre-test scores from 20-60%) that similar structured courses tend to produce similar gains in high-school, algebra-based, and calculus-based physics classes. We have chosen to compare to our own results rather than to the full Hake set since the Hake contained self-reported results from many classes and may have mixed together classes with distinct styles of reform. For our own classes, we know exactly what was done in each case.

[66] T. McCaskey, M. Dancy, and A. Elby, "Effects on assessment caused by splits between belief and understanding," in *2003 Physics Education Research Conference*, J. Marx, K. Cummings, and S. Franklin, eds., *AIP Conf. Proc.* **720**, 37-40 (2004).
19

# Reinventing College Physics for Biologists: Explicating an epistemological curriculum

E. F. Redish and D. Hammer

**Auxiliary Appendix: Supplementary Materials**

## 1. Epistemological Icons

| | |
|---|---|
| Restricting the scope of investigation | 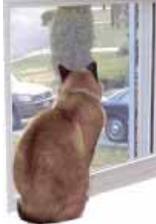 |
| Shopping for ideas | 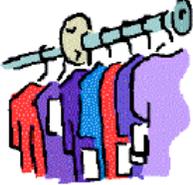 |
| Choosing foothold ideas | 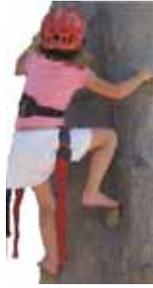 |
| The implications game | 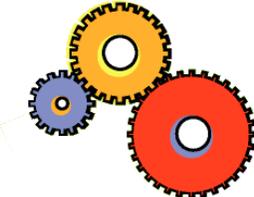 |
| Seeking coherence / Safety net | 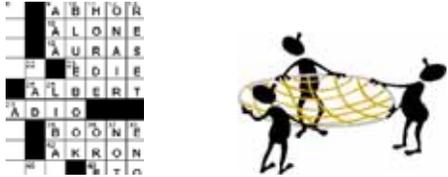 |
| Sense making | 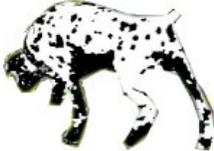 |
| Probing and refining intuitions | 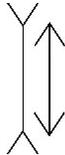 |
| Representation translation | 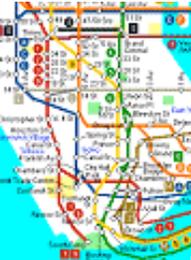 |

Explanations:
1. *Restricting the scope of investigation* – A critical idea in science is that we often describe only a limited set of phenomena. Trying to do everything at once doesn't let us make incremental progress. This restriction of scope is represented by a "cat television" icon. Cats love to sit and watch the world through the limited frame of a window.

2. *Shopping for ideas* – We use the metaphor of "shopping," with the clothing-rack icon, to help students think of their own knowledge and experience as having a large inventory of possibilities through which they could browse.

3. *Choosing foothold ideas* – We introduce the notion of a foothold idea as one we *choose* to accept as true, at least for the time being, as a way to proceed. As we find other ideas and findings fit with a foothold idea, and as we are able to respond to counter-arguments and counter-evidence, we form greater and greater commitment to the foothold; we are willing to work harder to reconcile other reasoning to fit with it.

4. *Playing the implications game* – Following the implications of our foothold ideas and principles is a critical process of science. Sometimes these reasoning's lead us to reject previously accepted principles. Sometimes they lead up to new phenomena. An engaging example is to use simple ideas about light to infer that a convex mirror should lead to non-existent objects appearing to float in mid-air. Students often reject this out of hand and don't realize that they would have then to reject their principles as well. The demonstration of the imaginary light bulb[1] is quite striking.

5. *Seeking coherence / safety net* – Building coherent pictures of an increasing set of diverse phenomena is a critical element of scientific thinking, not just for the purpose of confirming a theory, but for thinking out models and confirming ones thinking and intuition. We use one of two icons here, depending on whether we are focusing on coherence (the crossword puzzle) or self-checking (the safety net).

6. *Sense making* – Students often have a lot of trouble with elements of physics, especially equations. They tend to memorize their form without making sense of the relationships that the equations represent. We use the image shown in the figure below to illustrate this.

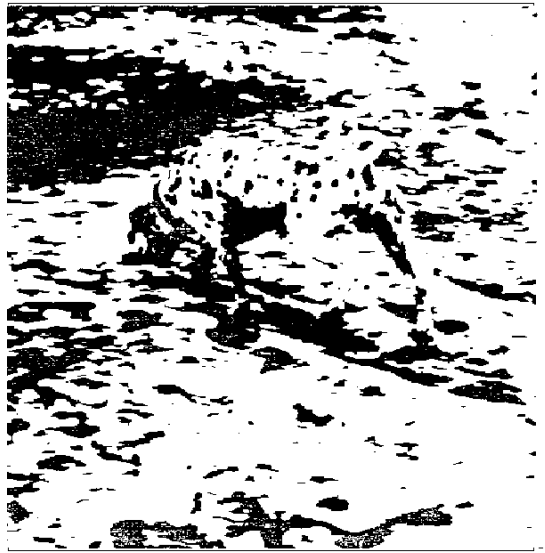

*Fig. S1)  A picture – in spots – of a Dalmatian dog drinking from puddles in the shadow of a leafy tree.*[2]

Many individuals see this only as a pattern of spots and can't pull together a coherent image. When students memorize an equation they are memorizing the spots rather than seeing the dog.

7. *Probing and refining intuitions* – This figure is known as the Lyell-Müller illusion in psychology. Both lines have the same length, but the arrows lead the eye, causing the viewer to interpret the line with the open arrowheads (>–<) as longer. This is used to remind the student that we cannot trust our first intuitions and often have to confirm our expectations via measurement.

8. *Representation translation* – We use a wide variety of ways of representing information about physical systems – diagrams, numbers, graphs, equations, and specialize figures. Different ways of representing the same physical system often help us understand the nature of the system. The pictures in fig. (S2) show three ways of representing the NYC subway system. The subway entrances are there in the highly accurate Landsat photo on the left, but it's not useful. The central schematic map is to scale, and shows the train lines and stops magnified and everything else ignored. But the lines tangle together in the lower end of Manhattan and are hard to separate. The representation on the right stretches the city so that the lines are easier to identify but it does not correctly represent relative distances. We use a piece of one of these maps when we are discussing representation of information in different ways.

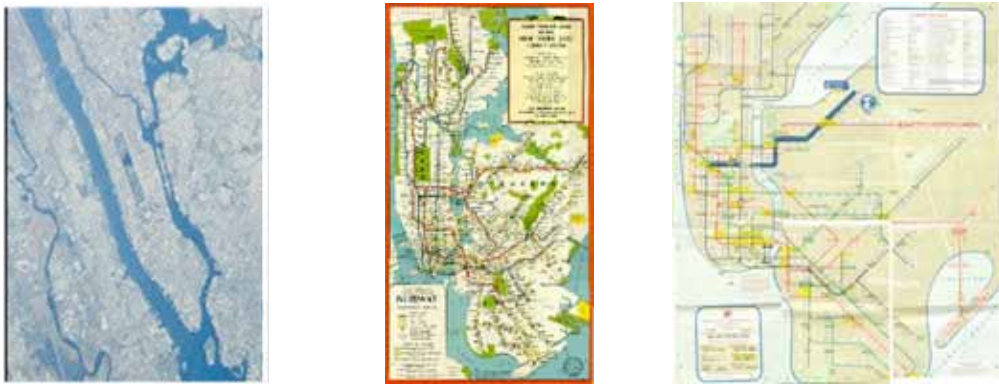

*Fig. S2)  Three representations of the New York City subway system*

## 2. Sample Problems

More problems of this type are available on the UMdPERG website at
http://www.physics.umd.edu/perg/abp/TPProbs/ProbSubjs.htm.

Figs. (S3) and (S4) display a lecture demonstration and a clicker question generated "on-the-fly" by one of us (DMH) in response to student questions.

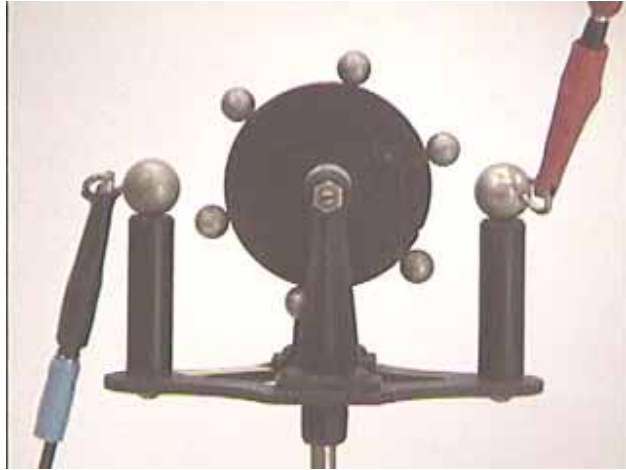

*Fig. S3) A: The electrostatic motor. When the potential produced by a Van de Graaff generator is applied across the terminals of the electrostatic motor as shown in the picture, the rotor of the electrostatic motor spins. As one of the six metal spheres on the rotor moves past a terminal it is charged by a discharge from the terminal. When it gets to the other side it is pulled toward the terminal, where it is charged to the opposite polarity and gets pushed away from the terminal. The motor can rotate either direction depending on initial conditions. [Courtesy R. Berg]$^3$*

> Could it turn the other way?
>   1. Yes
>   2. No
>
> If you disconnect the ground wire, will it still run?
>   1. Yes
>   2. No

*Fig. SS3)B: An "on-the-fly" pair of clicker questions generated in response to student questions about the electrostatic motor shown in figure SS3)A. The clicker questions led to vigorous discussions.*

Figure (SS4) shows a pre-prepared clicker problem presented by one of us (EFR) at the beginning of the first class on the motion of extended objects and before the presentation of the torque balance rule. It begins with a class discussion in which students are asked to comment on their personal experiences with balance, levers, and seesaws. The answers indicated 2-5 are drawn from the class. The students' intuitive suggestions (the man on the right holds up 5/6 of the cat's weight) leads, through a guided class discussion, to the torque balance rule. The problem is followed with a standard rigorous dem-

onstration of the rule. The emphasis throughout is on how we generate and test foothold ideas, following a path from intuition to hypothesis to experimental tests.

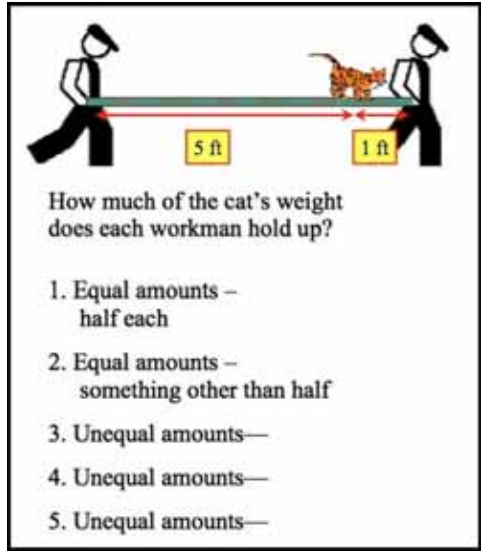

Fig. S4) : An example of an "epistemologized" pre-prepared Peer Instruction problem. For this and other PI problems, much of the "epistemologization" comes through guided class discussion.

Problems (S5) - (S9) show typical homework, exam, and quiz problems.

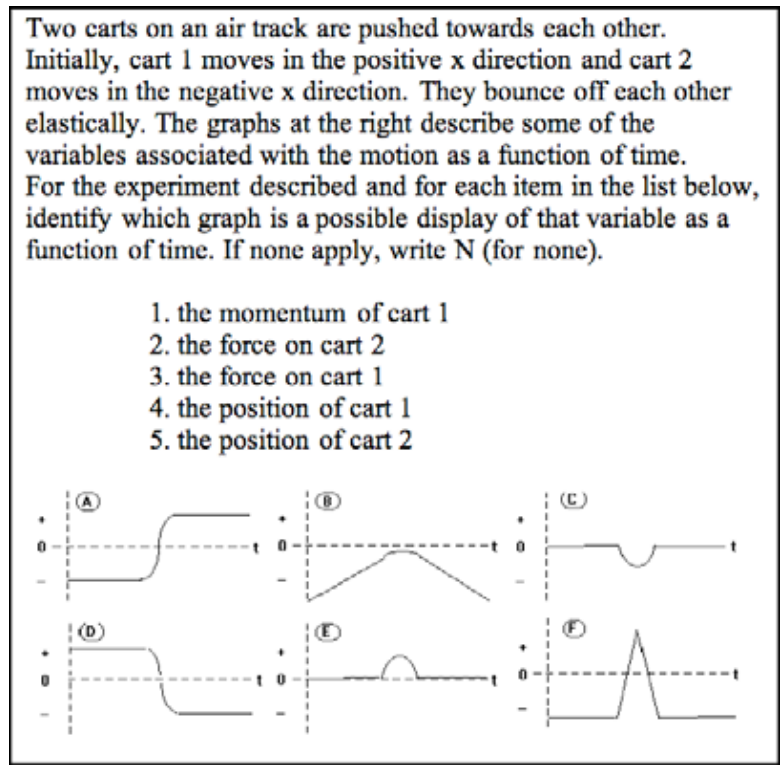

Fig. S5) A representation translation problem. When problems like this are given for homework, the students have to give written reasons for their answers. When they are given on exams, they do not.

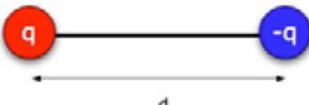

*Fig. S6) A "serious" homework problem that requires mixing different epistemological approaches, including bringing in personal experience, qualitative reasoning, estimation, and use of formal knowledge. Such a problem would not be given on an exam due to its length.*

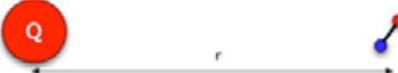

*Fig. S7) An essay question from a midterm hour exam.*

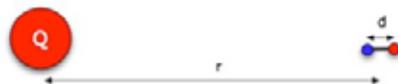

*Fig. S8) An estimation question from a midterm hour exam.*

1. You are sitting in a chair looking at two objects that are suspended from the ceiling. It appears to you that object A is above object B. When you stand up, object A appears to be below object B. Which of the two objects is farther away from you? (2 pts)

   a. Object A
   b. Object B
   c. They are both the same distance.
   d. You can't tell. It could be either one.

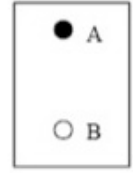
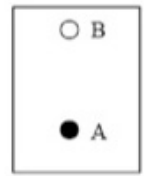

What you see while sitting    What you see while standing

*Fig. S9) A typical item in a weekly quiz. The quiz this week contained two items.*

## 3. MPEX-II Survey

Here are 25 statements (Items 1-25), which may or may not describe your beliefs about this course.
You are asked to rate each statement by selecting a response between A and E where the letters mean the following:

| A: Strongly Disagree | B: Disagree | C: Neutral | D: Agree | E: Strongly Agree |

Answer the questions by filling in the bubble on the scantron for the letter that best expresses your feeling. Work quickly. Don't over-elaborate the meaning of each statement. They are meant to be taken as straight-forward and simple.

<u>If you do not understand a statement, leave it blank. If you understand, but have no strong opinion one way or the other, choose C. If an item combines two statements and you disagree with either one, choose A or B.</u>

1.) Learning physics will help me understand situations in my everyday life.

2.) All I need to do to understand most of the basic ideas in this course is just go to lecture, work most of the problems, read the text, and/or pay close attention in class.

3.) The main point of seeing where a formula comes from is to learn that the formula is valid and that it is OK to use it in problems.

4.) When learning a new physics topic it's important to think about my personal experiences or ideas and relate them to the topic being analyzed.

5.) In this course, adept use of formulas is the main thing needed to solve physics problems effectively.

6.) Knowledge in physics consists of many pieces of information, each of which applies primarily to a specific situation.

7.) If I don't remember a particular equation needed for a problem in an exam I can probably figure out an (ethical!) way to come up with it, given enough time.

8.) Physics is related to the real world, but I can understand physics without thinking about that connection.

9.) "Problem solving" in physics basically means matching problems with facts or equations and then substituting values to get a number.

10.) In this course, I do not expect to understand equations in an intuitive sense; they just have to be taken as givens.

11.) When doing practice problems for a test or working on homework, if I came up with two different approaches to a problem and they gave different answers, I would not worry about it; after finding out the right answer, I'd just be sure to avoid the incorrect approach.

12.) My grade in this course will be primarily determined by how familiar I am with the material. Insight or creativity will have little to do with it.

13.) Often, a physics principle or theory just doesn't make sense. In those cases, you have to accept it and move on, because not everything in physics is supposed to make sense.

14.) If a problem on an exam does not look like one I've already done, I don't think I would have much of a chance of being able to work it out.

15.) Tamara just read something in her physics textbook that seems to disagree with her own experiences. But to learn physics well, Tamara shouldn't think about her own experiences; she should just focus on what the book says.

16.) The most crucial thing in solving a physics problem is finding the right equation to use.

17.) When handing in a physics test, you can generally have a correct sense of how well you did even before talking about it with other students.

18.) To really help us learn physics, professors in lecture should show us how to solve lots of problems, instead of spending so much time on concepts, proofs of general equations, and one or two problems.

19.) A significant problem in this course will be being able to memorize all the information I need to know.

20.) If physics professors gave really clear lectures with plenty of real-life examples and sample problems, then most good students could learn those subjects without having to spend a lot of time thinking outside of class.

21.) Although physical laws may apply to certain simple situations like we see in class and lab, they have little relation to what I experience in the real world.

22.) Group work in physics is beneficial only if at least one person in the group already understands and knows what they are talking about.

23.) When solving problems, the key thing is knowing the methods for addressing each particular type of question. Understanding the "big ideas" might be helpful for specially-written essay questions, but not for regular physics problems.

24.) To understand physics, the formulas (equations) are really the main thing; the other material is mostly to help you decide which equations to use in which situations.

25.) It wouldn't matter if I didn't get my homework returned to me as long as I knew which questions I got wrong and I had the solutions to study.

26.) Two students are talking about their experiences in class:

   Meena: Our group is really good, I think. We often spend a lot of time confused and sometimes never feel like we have the right answer, but we all listen to each other's ideas and try to figure things out that way.

   Salehah: In our group there is one person who always knows the right answer and so we pretty much follow her lead all the time. This is a great because we always get the tasks done on time and sometimes early.

  (a)   I agree almost entirely with Meena.
  (b)   Although I agree more with Meena I think Salehah makes some good points.
  (c)   I agree (or disagree) equally with Meena and Salehah.
  (d)   Although I agree more with Salehah, I think Meena makes some good points.
  (e)   I agree almost entirely with Salehah.

27.) In the following question, you will read a short discussion between two students who disagree about some issue. Then you'll indicate whether you agree with one student or the other.

   Tracy: A good physics textbook should show how the material in one chapter relates to the material in other chapters. It shouldn't treat each topic as a separate "unit," because they're not really separate.

   Carissa: But most of the time, each chapter is about a different topic, and those different topics don't always have much to do with each other. The textbook should keep everything separate, instead of blending it all together.

   With whom do you agree? Read all the choices before choosing one.

(a) I agree almost entirely with Tracy.
(b) Although I agree more with Tracy, I think Carissa makes some good points.
(c) I agree (or disagree) equally with Carissa and Tracy.
(d) Although I agree more with Carissa, I think Tracy makes some good points.
(e) **I agree almost entirely with Carissa.**

28.) Let's say a student has limited time to study, and therefore must choose between the following options. Assuming the exam will be a fair test of understanding, and assuming time pressure during the exam isn't an issue, which option should the student choose?

(a) Learning only a few basic formulas, but going into depth with them.
(b) Learning all the formulas from the relevant chapters, but not going into as much depth.
(c) Compromising between (a) and (b), but leaning more towards (a).
(d) Compromising between (a) and (b), but leaning more towards (b).
(e) Compromising between (a) and (b), midway between those two extremes.

29.) Some people have 'photographic memory', the ability to recall essentially everything they read. To what extent would photographic memory give you an advantage when learning physics?

(a) It would be the most helpful thing that could happen to me
(b) It would help a lot
(c) It would help a fair amount
(d) It would help a little
(e) It would hardly help at all

30.) Consider the following question from a popular textbook:

"A horse is urged to pull a wagon. The horse refuses to try, citing Newton's 3rd law as a defense: The pull of the horse on the wagon is equal but opposite to the pull of the wagon on the horse. 'If I can never exert a greater force on the wagon than it exerts on me, how can I ever start the wagon moving?' asks the horse. How would you reply?"

When studying for a test, what best characterizes your attitude towards studying and answering questions such as this?

(a) Studying these kinds of questions isn't helpful, because they won't be on the test.
(b) Studying these kinds of questions helps a little bit, but not nearly as much studying other things (such as the problem-solving techniques or formulas).
(c) Studying these kinds of questions is fairly helpful, worth a fair amount of time.
(d) Studying these kinds of questions is quite helpful worth quite a lot of my time.
(e) Studying these kinds of questions is extremely helpful, worth a whole lot of my study time.

31.) Roy and Theo are working on a homework problem.

Roy: "I remember in the book it said that anything moving in a circle has to have a centripetal acceleration."
Theo: "But if the particle's velocity is constant, how can it be accelerating? That doesn't make sense."
Roy: "Look, right here, under 'Uniform Circular Motion' – here's the equation, $a=v^2/r$. That's what we need for this problem."
Theo: "But I know that to have an acceleration, we need a change in velocity. I don't see how the velocity is changing. That equation doesn't seem right to me."

If you could only work with one of them, who do you think would be more helpful?

- (a) Roy would be much more helpful.
- (b) Roy would be a little more helpful.
- (c) They would be equally helpful.
- (d) Theo would be a little more helpful.
- (e) Theo would be much more helpful.

32.) Several students are talking about group work.

Carmela: "I feel like explaining something to other people in my group really helps me understand it better."

Juanita: "I don't think explaining helps you understand better. It's just that when you can explain something to someone else, then you know you already understood it."

With whom do you agree? Read all the choices before choosing one.

- (a) I agree almost entirely with Carmela.
- (b) Although I agree more with Carmela, I think Juanita makes some good points.
- (c) I agree (or disagree) equally with Juanita and Carmela.
- (d) Although I agree more with Juanita, I think Carmela makes some good points.
- (e) I agree almost entirely with Juanita.

## MPEX-II Category groupings

The items are divided into clusters according to the intent of the researchers. Note these are not necessarily functionally independent. They are not intended to be orthogonal factors. The indented topics below the main categories of coherence, concepts, and independence are sub-categories.

*Coherence*: The extent to which the student sees physics knowledge as coherent and sensible as opposed to a bunch of disconnected pieces.
    3, 4, 6, 8, 10, 13, 15, 19, 21, 23, 27, 28
  *Coherence-math*: Coherence between math formalism
     and physics intuitions and concepts.  3, 10, 28
  *Coherence-reality*: Coherence between what's taught in the classroom and what's experienced in the real world.  4, 8, 15, 21
  *Coherence-other*: Everything else.  Similar to MPEX-I coherence.  6, 13, 19, 23, 27

*Concepts*: The extent to which students see concepts as the substance of physics -- as opposed to thinking of them as mere cues for which formulas to use.
    5, 9, 16, 18, 19, 23, 24, 28, 30

*Independence*: The extent to which the student sees learning physics as a matter of constructing her own understanding rather than absorbing knowledge from authority. Similar to MPEX-I independence.
    2, 7, 11, 12, 14, 17, 20, 22, 25, 29, 31, 32
  *Independence-epistemology*: The aspect of independence that relates to the student's view of the nature of the knowledge being learnd.  2, 11, 12, 20, 22, 25, 29, 31, 32
  *Independence-personal*: The self-efficacy the student feels about her ability to construct understanding as opposed to just accept what the instructor says. 7, 14, 17

## 4. CORE Survey

This is not a teaching evaluation. This course is approved as a Distributive Studies course in the CORE Liberal Arts and Science Studies Program, the campus general education program. Distributive Studies courses are intended to provide opportunities to learn about the fundamental ideas and issues central to a major intellectual discipline. Stated CORE goals include active learning, critical thinking, and writing. Your responses to the following questions will help to show us to what extent you think this course met these goals. We value your help in this endeavor

Instructions:

Use a #2 pencil. Please enter your responses on the machine-scannable answer sheet provided by darkening the appropriate spaces for your responses. Answer each item on a scale of (1) to (5) where the numbers mean:

(1) not at all   (2) a little   (3) somewhat   (4) quite a lot   (5) a great deal

1. To what extent has this course made you aware of a collection of ideas, theories or concepts that are central to this field?
2. To what extent has this course helped you understand the ways experts in this field think?
3. To what extent has this course helped you understand the method of study, or observation, collection, and analysis of characteristic of this field?
4. To what extent have the writing assignments and/or examinations in this course given you opportunities to think carefully and critically?
5. To what extent has this course given you opportunities to participate actively in the learning process through discussions, small group work, laboratories, etc.?
6. To what extent has this course been intellectually stimulating?
7. To what extent has the syllabus for this course been an accurate guide to what has happened in class and what has been expected of you?
8. To what extent has this course challenged you to examine your knowledge and experience in new ways and/or explore new ideas and ways of thinking?

[1] R. E. Berg, Physics Lecture Demonstrations, http://www.physics.umd.edu/deptinfo/facilities/lecdem/services/demos/demosl3/l3-21.htm

[2] John P. Frisby, *Seeing: Illusion, Brain and Mind* (Oxford University Press, 1980).

[3] R. E. Berg, Physics Lecture Demonstrations, http://www.physics.umd.edu/lecdem/services/demos/demosj2/j2-51.htm.